\documentclass[9pt,twocolumn,twoside]{opticajnl}
\journal{opticajournal} % use for journal or Optica Open submissions

\setboolean{shortarticle}{true}
\title{Synthetic-Aperture Super-Resolution Imaging via Spatial-Frequency Shift in Near-Field Diffraction}

\author[1]{Qihao Sun}
\author[1]{Chunzheng Bai}
\author[1]{Wenjing Fang}
\author[2,*]{Zongyan Zhang}
\author[3]{Songlin Yang}
\author[3]{Yonghong Ye}
\author[4]{Mingyi Tao}
\author[1,*]{Jiayu Zhang}

\affil[1]{School of Electronic Science and Engineering, Southeast University, Nanjing 211189, China}
\affil[2]{School of Optoelectronic Engineering, Henan Normal University, Xinxiang 453007, China}
\affil[3]{School of Computer and Electronic Information, Nanjing Normal University, Nanjing 210023, China}
\affil[4]{Yangzhou Jinghong Laser Technology Co., Ltd., Yangzhou 225101, China}

\affil[*]{Corresponding authors: zhangzongyan@htu.edu.cn; jyzhang@seu.edu.cn}

\begin{abstract}
Grating-based computational imaging shifts high-spatial-frequency information into the detectable range, enabling super-resolution reconstruction. However, a fixed grating produces effective diffraction-mediated shifts only for specific spatial-frequency vectors, limiting spatial-frequency coverage. Here, we propose a synthetic-aperture super-resolution method in which repeated imaging with a rotating grating broadens spatial-frequency coverage and expands the effective aperture. A physics-prior-guided restoration framework then employs a multichannel deep-learning network to fuse images acquired at different grating orientations. Using diffraction images from only three grating orientations, the proposed method reconstructed the object with a resolution of $\lambda/3.9$. This approach offers a practical route to grating-modulated super-resolution imaging in systems with limited numerical aperture.
\end{abstract}

\setboolean{displaycopyright}{false} % Do not include copyright or licensing information in submission.

\begin{document}

\maketitle

Optical diffraction generally limits the spatial resolution of conventional microscopes to approximately half the wavelength of light. Although recent advances in fluorescence super-resolution microscopy have enabled nanoscale imaging, these techniques typically require specific fluorescent labels~\cite{reinhardtAngstromresolutionFluorescenceMicroscopy2023}. In contrast, label-free super-resolution imaging~\cite{jayakumarChipbasedLabelfreeIncoherent2025} directly captures the intrinsic structural and optical information of a sample, offering substantial value for materials characterization, nanostructure inspection, and biological imaging. A range of approaches has been developed, including near-field scanning optical microscopy~\cite{sidayAllopticalSubcycleMicroscopy2024}, superlenses and hyperlenses~\cite{fangSubDiffractionLimitedOpticalImaging2005,liuFarFieldOpticalHyperlens2007}, and microsphere-assisted super-resolution microscopy~\cite{parkMicrosphereassistedHyperspectralImaging2024}. These methods enable label-free imaging of subwavelength structures by locally detecting evanescent fields carrying high-spatial-frequency information, enhancing their transmission, or converting them into propagating waves. Another widely studied strategy is the spatial-frequency shift (SFS)~\cite{pangSpatialfrequencyshiftEnablesIntegrated2022}, in which structured illumination or near-field modulation structures shift part of the high-frequency information originally beyond the system's cutoff spatial frequency into the detectable spatial-frequency band of the objective. Structured illumination microscopy~\cite{watanabeStructuredLineIllumination2015}, Fourier ptychographic microscopy~\cite{zhengWidefieldHighresolutionFourier2013}, and synthetic-aperture microscopy based on multi-angle illumination~\cite{tamamitsuMidinfraredWidefieldNanoscopy2024} all exploit this concept of spectral shifting and synthesis. By combining multiple acquisitions under different illumination conditions, these methods extend the effective spatial-frequency coverage and improve resolution.

Near-field diffraction can convert evanescent waves carrying high-spatial-frequency information about the sample surface into waves that propagate to the far field. Previous studies have explored several routes to grating-assisted super-resolution imaging, including the use of proximity projection gratings to generate fine structured illumination~\cite{huSub100NmResolution2015}, one- and two-dimensional Fibonacci quasiperiodic gratings to convert sample evanescent waves into propagating waves~\cite{wuTwodimensionalFibonacciGrating2016,wuOnedimensionalFibonacciGrating2013a}, and gratings with different periods and orientations to tune the illumination wavevector~\cite{tangHighRefractiveIndexChipPeriodically2022a}. Nevertheless, grating-assisted super-resolution imaging still faces two challenges. First, after grating diffraction, information carried by different diffraction orders overlaps within the detection band, making high-frequency information more difficult to separate and recover~\cite{liuFarFieldOpticalSuperlens2007a}. Previous reconstruction methods have included spectral separation, shifting, and stitching~\cite{liuExperimentalStudiesFarfield2007}, as well as regularized deconvolution based on the system transfer function~\cite{wuTwodimensionalFibonacciGrating2016}. Recent studies suggest that deep learning offers a new route to reconstruction in grating-assisted and SFS-based super-resolution imaging. Learning the mapping between grating-modulated images and target structures, together with attention mechanisms or joint optimization in the spatial and frequency domains, can facilitate high-frequency information recovery and improve reconstruction efficiency and noise robustness~\cite{liuSuperresolutionMicroscopyGrating2024e,liuSpatialFrequencyShift2024b,zhangAlternativeDeepLearning2023a}. Second, a trade-off remains among spatial-frequency coverage, grating structural complexity, and data acquisition efficiency. The principal spatial-frequency shift direction of a fixed one-dimensional grating is determined by its orientation, limiting its ability to capture high-frequency information along other directions~\cite{liuSuperresolutionMicroscopyGrating2024e}. Although two-dimensional, quasiperiodic, or multilevel gratings can extend spatial-frequency coverage, they may increase design and fabrication complexity. Moreover, modulation in multiple directions or at multiple levels generally requires more raw images, limiting acquisition efficiency~\cite{tangHighRefractiveIndexChipPeriodically2022a}.

Here, we propose a synthetic-aperture super-resolution imaging method based on spatial-frequency shifts induced by near-field diffraction. Conventional synthetic-aperture methods acquire information from different regions of the frequency domain by varying the illumination angle or detection position. Our method instead actively controls the spatial-frequency shift direction by rotating a grating. Multiple grating orientations extend the overall spatial-frequency support of the system, forming an effective synthetic aperture. We further develop a spatial-frequency synthesis network (SFSyn-Net) guided by physical priors to jointly reconstruct a small number of diffraction images acquired at different grating orientations. This enables effective fusion of high-frequency information from different directions and recovery of subwavelength structures. Experimentally, acquisitions at only three grating angles achieve a spatial resolution of approximately $140~\mathrm{nm}$, or $\lambda/3.9$, under $540~\mathrm{nm}$ illumination. A simple grating and acquisitions at a few angles are sufficient to capture high-frequency information and expand the synthetic aperture, without requiring a complex multilevel grating design. This approach provides a new strategy for efficient and flexible label-free far-field super-resolution imaging under limited numerical aperture conditions.

\begin{figure}[t]
\centering
\includegraphics[width=\linewidth]{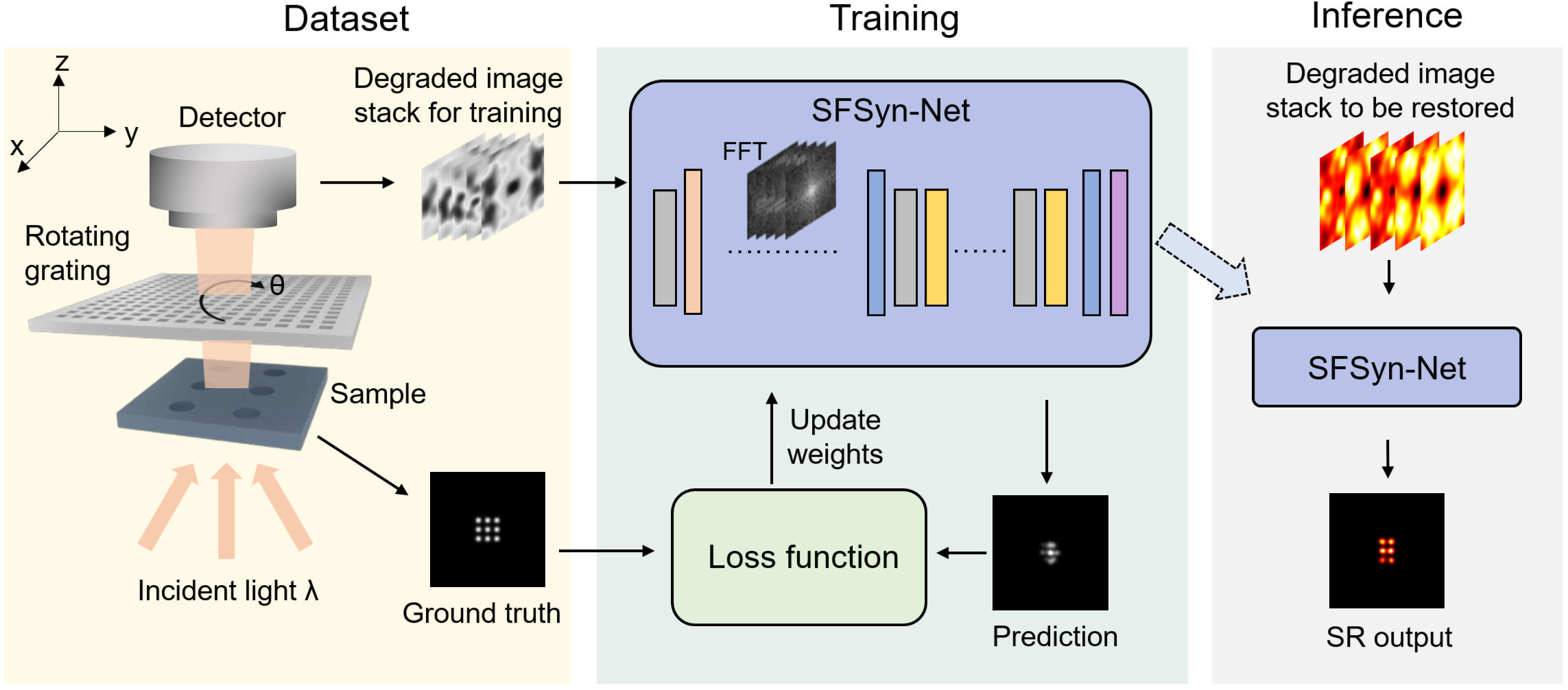}
\caption{Workflow of SFSyn-Net guided by physical priors, comprising dataset construction (left), network training (center), and inference (right). Multi-angle diffraction images generated by a rotating grating are fused under the guidance of a Fourier-domain spatial-frequency shift prior to reconstruct a super-resolution image.}
\label{fig1}
\end{figure}

The training and reconstruction workflow of the super-resolution imaging network is shown in Fig.~\ref{fig1} and consists of three stages: dataset construction, network training, and image inference. During dataset construction, the finite-difference time-domain (FDTD) method is used to establish a physical forward model of near-field diffraction imaging and generate data for network training and performance validation. By incorporating physical priors of the imaging process into data generation, paired diffraction observations and high-resolution structural labels can be obtained without experimental acquisition, reducing dependence on experimental data and measured high-resolution ground truth. The computational domain measures \(8~\mu\mathrm{m} \times 8~\mu\mathrm{m} \times 5~\mu\mathrm{m}\), with perfectly matched layers at its boundaries to absorb outgoing electromagnetic waves. The samples consist of arrays of circular apertures, each \(100~\mathrm{nm}\) in diameter, arranged in various two-dimensional configurations. Each sample is placed in close contact with the grating to enable effective coupling of its near-field evanescent components to the grating. The two-dimensional periodic grating has a period of \(160~\mathrm{nm}\) in both the \(x\) and \(y\) directions, with apertures measuring \(110~\mathrm{nm} \times 110~\mathrm{nm}\) and a \(50~\mathrm{nm}\) gap between adjacent apertures. Incident light at a wavelength of \(700~\mathrm{nm}\) passes through the sample and then the grating. Diffracted light carrying high-frequency structural information and transmitted light carrying low-frequency information are collected and recorded in the far field to produce a degraded image of the sample. By varying the grating orientation angle \(\theta\), five degraded images are acquired for each sample and stacked in orientation order to form a multichannel input. The corresponding sample structure image serves as the ground-truth label. 

\begin{figure}[t]
\centering
\includegraphics[width=0.6\linewidth]{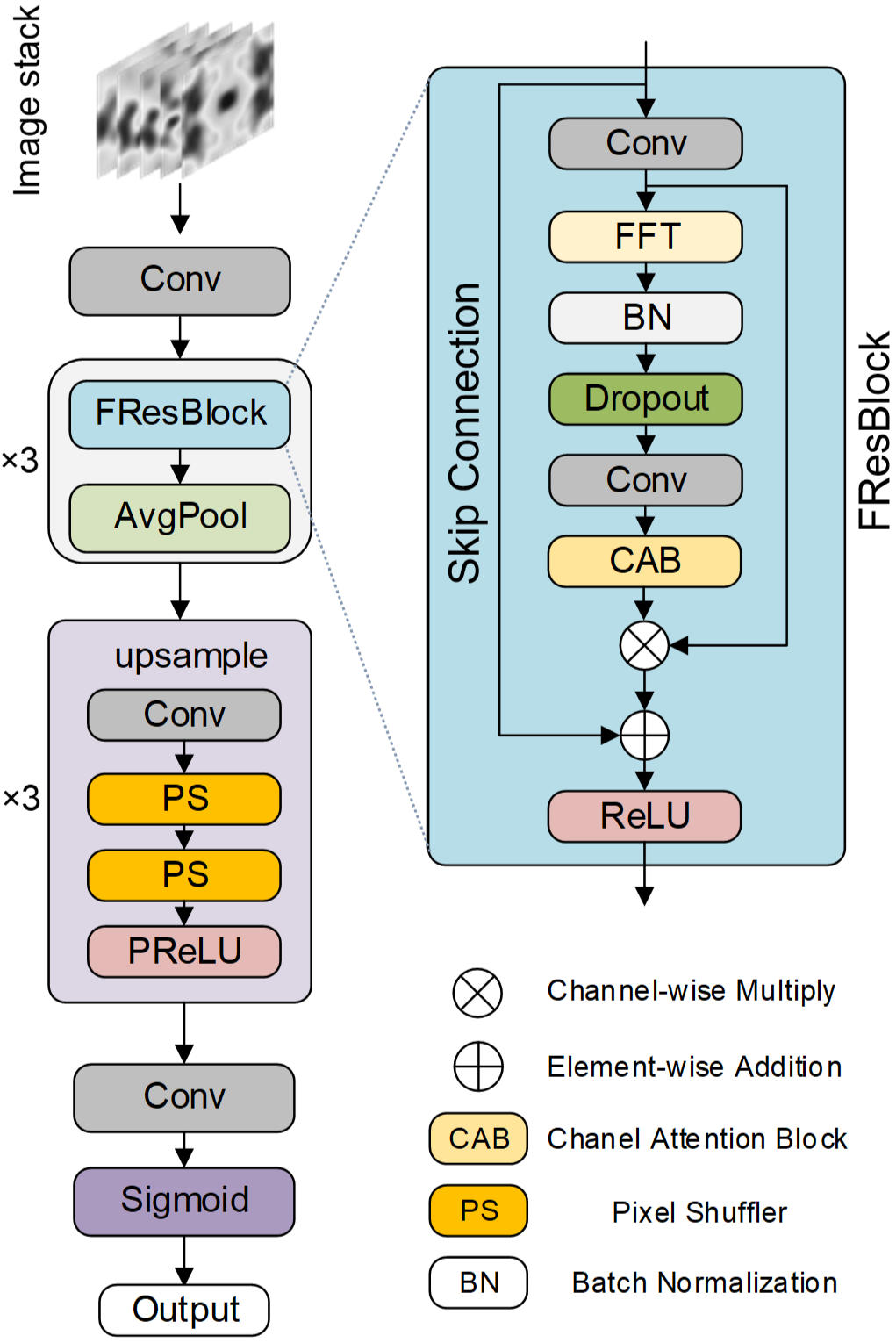}
\caption{Architectures of SFSyn-Net and its FResBlock. The multichannel image stack passes sequentially through three FResBlock–average pooling units, three pixel-shuffle-based upsampling units, and an output layer.}
\label{fig2}
\end{figure}

During network training, degraded images acquired at different grating orientations are stacked in angular order as a multichannel input. The spatial-frequency synthesis network (SFSyn-Net) learns the mapping between diffraction observations and high-resolution sample structures. As shown in Fig.~\ref{fig2}, a convolutional layer first maps the input to shallow features~\cite{dongImageSuperResolutionUsing2016}, followed by three FResBlock--average-pooling units that extract multiscale features. FResBlock combines spatial features with a spectrally guided branch, using a Fourier transform and channel attention to generate adaptive weights for weighted enhancement of the spatial features~\cite{qiaoEvaluationDevelopmentDeep2021}, while a residual connection preserves the input information~\cite{zhangResidualDenseNetwork2021}. The features then pass through three pixel-shuffle-based upsampling modules to restore spatial resolution. A convolutional layer followed by a sigmoid activation function produces the reconstructed image. The discrepancy between the predicted image and the ground-truth label is evaluated using a hybrid loss function that combines the structural similarity index (SSIM) and the \(L_1\) norm (SSIM \(+ L_1\))~\cite{jadhavArtefactRemovalGround2021}. The SSIM term constrains structural consistency in the reconstruction, whereas the \(L_1\) term constrains pixel-wise error. Network weights are iteratively updated through backpropagation using the Adam optimizer, enabling the network to learn from multi-angle observations and recover fine structural details of the sample.

Finally, during image inference, the trained network weights are fixed. The multi-angle degraded images to be restored are stacked in the prescribed order and fed into SFSyn-Net, which produces the corresponding super-resolution reconstruction in a single forward pass.

\begin{figure}[t]
\centering
\includegraphics[width=0.8\linewidth]{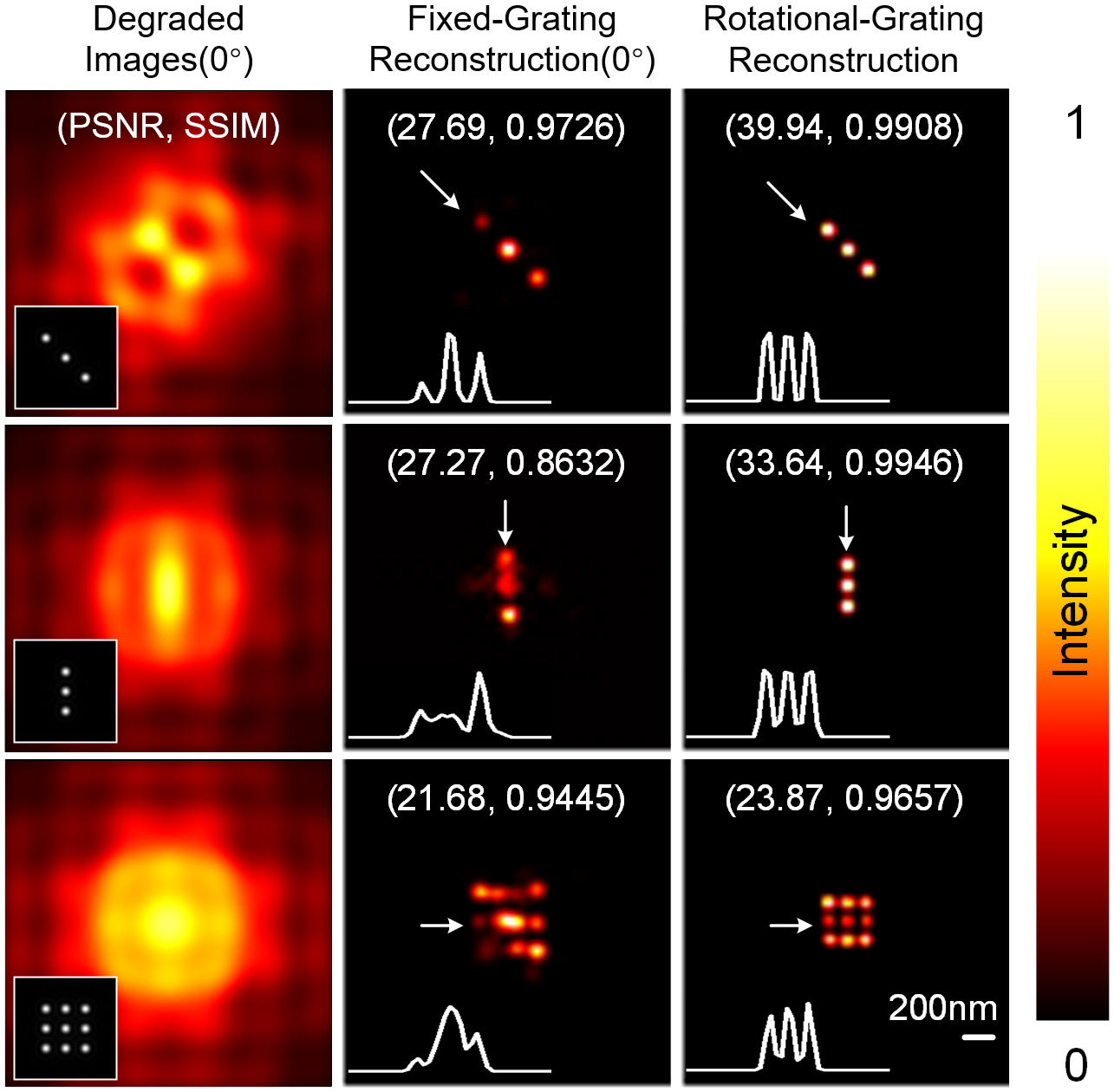}
\caption{Reconstructions of subwavelength arrays with a center-to-center spacing of 160 nm. From top to bottom: an oblique three-point array, a vertical three-point array, and a 3 × 3 array. From left to right: degraded images at 0°, fixed-grating reconstructions at 0°, and rotating-grating synthetic-aperture reconstructions. For fixed-grating reconstruction, all five degraded images in the input stack are identical. The inset for the first example shows the ground-truth target structure. White curves show normalized intensity profiles at the positions indicated by the arrows. Values in parentheses denote the peak signal-to-noise ratio (PSNR, in dB) and SSIM, respectively.}
\label{fig3}
\end{figure}

To evaluate the inference performance of SFSyn-Net, three sample structures excluded from the training set were selected, as shown in Fig.~\ref{fig3}. A physical forward model of near-field grating diffraction was used to generate degraded image stacks, which were then restored using SFSyn-Net. The simulation results demonstrate the ability of the rotating-grating synthetic-aperture method to recover subwavelength structures, achieving a resolution of \(160~\mathrm{nm}\), approximately \(\lambda/4.4\). This is comparable to the resolution previously achieved using quasiperiodic gratings with complex structures~\cite{liuSpatialFrequencyShift2024b}.

Because of the diffraction limit, conventional microscopy cannot resolve adjacent point structures separated by less than \(\lambda/2\), and all three targets appear as blurred spots. Fixed-grating reconstruction recovers some structural information, but the limited spatial-frequency coverage provided by a single grating orientation leads to nonuniform spot intensities, incomplete separation of neighboring structures, and distorted array morphology. By acquiring complementary spatial-frequency information at different orientations, the rotating grating more completely recovers the geometry of all three targets. Clear intensity valleys separate adjacent points, and three distinct peaks can be identified in the intensity profiles. For the three targets, the rotating-grating method increases PSNR from \(27.69\), \(27.27\), and \(21.68~\mathrm{dB}\) to \(39.94\), \(33.64\), and \(23.87~\mathrm{dB}\), respectively, and SSIM from \(0.9726\), \(0.8632\), and \(0.9445\) to \(0.9908\), \(0.9946\), and \(0.9657\), respectively. These results indicate that rotation through multiple angles alleviates the insufficient spatial-frequency coverage of a single grating orientation, improving the completeness and accuracy of subwavelength structure reconstruction.

\begin{figure}[t]
\centering
\includegraphics[width=\linewidth]{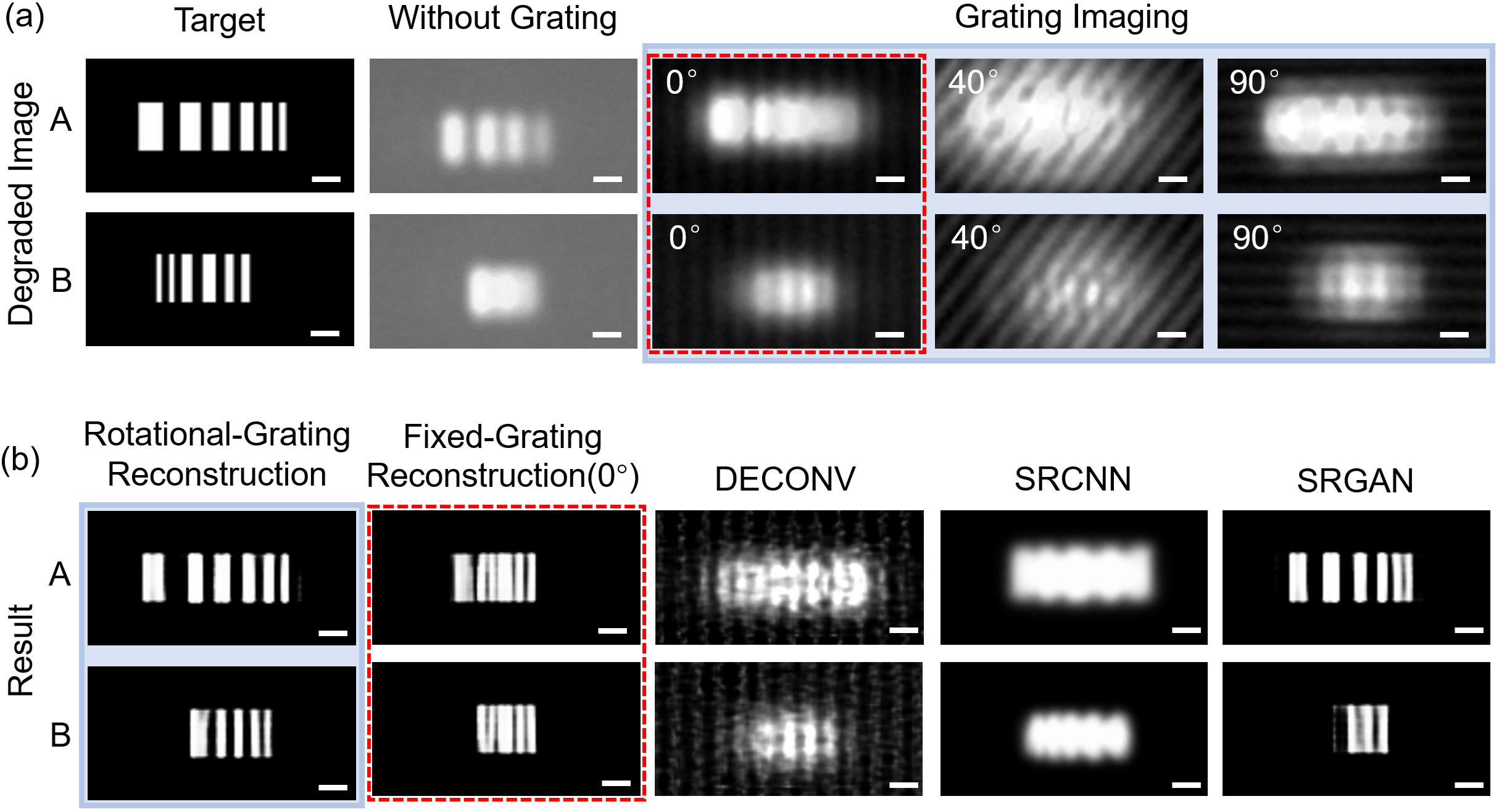}
\caption{Comparison of reconstruction methods for two line-array targets, A and B. (a) Target structures and degraded images acquired without a grating and at grating orientations of 0°, 40°, and 90°. (b) Results obtained using rotating-grating reconstruction, fixed-grating reconstruction at 0°, deconvolution (DECONV), SRCNN, and SRGAN. Blue solid boxes and red dashed boxes indicate multi-angle and single-angle data and their corresponding results, respectively. From left to right, the nominal stripe widths are 950, 800, 650, 500, 400, and 250 nm for sample A, and 200, 200, 400, 500, 300, and 300 nm for sample B.}
\label{fig4}
\end{figure}

In the experiments, the sample was illuminated by an LED source at a wavelength of \(540~\mathrm{nm}\). A one-dimensional grating with a period of \(900~\mathrm{nm}\) and a groove width of \(600~\mathrm{nm}\) was placed between the sample and the microscope objective. After modulation by the grating, the outgoing optical field carrying the sample's structural information was collected by an objective with a numerical aperture (NA) of \(0.75\) and relayed to a CMOS camera. We further augmented the dataset with training samples containing structures such as lines and stripes and trained SFSyn-Net on the expanded dataset. Images were acquired sequentially at grating orientations of \(0^\circ\), \(40^\circ\), and \(90^\circ\). Figure~\ref{fig4}(a) shows the designed structures of the two samples, the microscopy images acquired without a grating, and the raw observations acquired at different grating orientations. Imaging without a grating preserves the overall intensity profile of each sample, but diffraction blur causes substantial overlap between neighboring stripes. With the grating introduced, the raw images exhibit distinctly angle-dependent diffraction patterns. As shown in Fig.~\ref{fig4}(b), fixed-grating reconstruction using only the \(0^\circ\) image recovers some stripe structures but still exhibits distorted stripe widths, nonuniform intensities, and missing local structures. We also compared image restoration using DECONV~\cite{richardsonBayesianBasedIterativeMethod1972}, SRCNN~\cite{dongImageSuperResolutionUsing2016a}, and SRGAN~\cite{ledigPhotoRealisticSingleImage2017}. Deconvolution amplifies noise and introduces periodic artifacts, making the sample structures difficult to identify accurately. SRCNN produces smoother outputs, but adjacent stripes remain merged and the recovery of high-frequency details is limited. SRGAN produces sharper edges, yet the number, widths, and spatial positions of some stripes differ from those in the target image, indicating that improved visual sharpness does not necessarily imply more accurate structural recovery. By fusing complementary information from different grating angles, SFSyn-Net more completely recovers the arrangement orientations and relative stripe widths of both samples while introducing fewer background artifacts. Its reconstructions most closely resemble the target images.

The PSNR and SSIM values in Table~\ref{tab1} are broadly consistent with the visual comparison in Fig.~\ref{fig4}. For samples A and B, the proposed method achieves PSNR values of \(15.34\) and \(13.77~\mathrm{dB}\) and SSIM values of \(0.8594\) and \(0.8338\), respectively, exceeding those of deconvolution, SRCNN, and SRGAN. Relative to SRGAN, the best-performing comparison method, the proposed method improves PSNR by \(3.19\) and \(1.97~\mathrm{dB}\) and SSIM by \(0.0987\) and \(0.1281\) for the two samples, respectively. On the present experimental dataset and under the same evaluation conditions, the proposed method achieves higher structural fidelity and more complete recovery of stripe details.

Finally, to quantitatively assess the lateral resolving capability of the reconstructed images, a rectangular function representing the stripe structure was convolved with a Gaussian response function and fitted to the experimental stripe intensity profiles~\cite{batesMulticolorSuperResolutionImaging2007}. The procedure is detailed in Supplementary Fig.~S1. Using the full width at half maximum (FWHM) of the fitted Gaussian response to characterize effective resolution, direct imaging without a grating and rotating-grating reconstruction yield resolutions of \((453.8 \pm 7.3)~\mathrm{nm}\) and \((140 \pm 8.7)~\mathrm{nm}\), respectively. The latter is finer than the resolution previously demonstrated experimentally with a fixed one-dimensional grating, approximately \(\lambda/3\)~\cite{liuSuperresolutionMicroscopyGrating2024e}.

\begin{table}[htbp]
\centering
\caption{\bf Comparison of restoration quality for different methods.}
\label{tab1}

\resizebox{\columnwidth}{!}{%
\begin{tabular}{lcccc}
\hline
& \multicolumn{2}{c}{\textbf{Sample A}}
& \multicolumn{2}{c}{\textbf{Sample B}} \\
\cline{2-5}
& \textbf{PSNR (dB)} & \textbf{SSIM}
& \textbf{PSNR (dB)} & \textbf{SSIM} \\
\hline
Degraded image & 6.12  & 0.0614 & 5.81  & 0.0544 \\
DECONV         & 10.54 & 0.0721 & 10.62 & 0.0352 \\
SRCNN          & 9.99  & 0.5329 & 11.57 & 0.6926 \\
SRGAN          & 12.15 & 0.7607 & 11.80 & 0.7057 \\
Ours           & 15.34 & 0.8594 & 13.77 & 0.8338 \\
\hline
\end{tabular}%
}
\end{table}

In summary, we have demonstrated a far-field super-resolution imaging method that combines near-field grating modulation with a spatial-frequency synthesis network. The method features a simple optical configuration, requires observations at only a few angles, and can be readily integrated with conventional microscopes. Physical priors of the imaging process are incorporated into training data construction, with numerical simulations generating paired diffraction observations and structural labels. This reduces reliance on experimental acquisition and measured high-resolution ground truth and facilitates the expansion of training data for different sample types. Experimentally, at a wavelength of \(540~\mathrm{nm}\), the estimated effective resolution improves from \((453.8 \pm 7.3)~\mathrm{nm}\) without a grating to \((140 \pm 8.7)~\mathrm{nm}\), approximately \(\lambda/3.9\). Future work could extend the simulated samples to nonperiodic structures and local defects and account for grating orientation errors, variations in sample--grating separation, and experimental noise. These extensions could improve network generalization and reconstruction robustness and broaden applications in micro- and nanostructure characterization and defect detection. We anticipate that this method could serve as an add-on imaging module for conventional microscopes and, through integration with multi-angle synthetic-aperture imaging, provide a structurally simple and readily integrable optical characterization approach for far-field super-resolution microscopy.

\begin{backmatter}
\bmsection{Funding} National Natural Science Foundation of China (Grant No. 62375135).

\bmsection{Acknowledgment} The section title should not follow the numbering scheme of the body of the paper. Additional information crediting individuals who contributed to the work being reported, clarifying who received funding from a particular source, or other information that does not fit the criteria for the funding block may also be included; for example, ``K. Flockhart thanks the National Science Foundation for help identifying collaborators for this work.''

\bmsection{Disclosures} The authors declare no conflicts of interest.

\bmsection{Data Availability Statement} All data generated or analyzed during this study are included in this article and its Supplementary Material. The underlying raw data are available from the corresponding author upon reasonable request.

\bmsection{Supplemental document}
See Supplementary Material for supporting content, which presents the estimation of the effective lateral resolution from stripe-edge responses.
\end{backmatter}

% Bibliography
\bibliography{sample}

% Full bibliography added automatically for Optics Letters submissions; the following line will simply be ignored if submitting to other journals.
% Note that this extra page will not count against page length
%\bibliographyfullrefs{sample}

%Manual citation list
%\begin{thebibliography}{1}
%\bibitem{Zhang:14}
%Y.~Zhang, S.~Qiao, L.~Sun, Q.~W. Shi, W.~Huang, %L.~Li, and Z.~Yang,
 % \enquote{Photoinduced active terahertz metamaterials with nanostructured
  %vanadium dioxide film deposited by sol-gel method,} Opt. Express \textbf{22},
  %11070--11078 (2014).
%\end{thebibliography}

\end{document}

\begin{center}
{\LARGE\bfseries\color{sectionblue} Supplementary Material}
\end{center}

\section{Estimation of effective lateral resolution}

This section describes the estimation of the effective lateral resolution
from stripe-edge responses for direct imaging without a grating and
rotating-grating reconstruction.

Figures S1(a) and S1(b) show direct imaging without a grating and
rotating-grating reconstruction, respectively. ROI1 and ROI2 denote the edge
regions used for analysis. In Figs. S1(c) and S1(f), each region of interest
(ROI) is divided into eight horizontal bands, and the red dots indicate the
fitted edge centers. Figures S1(d) and S1(g) show the edge spread functions
(ESFs) extracted from the individual bands and their fits. Each ESF is fitted
using the convolution of an ideal step-edge function with a Gaussian point
spread function. The fitted Gaussian standard deviation $\sigma$ is converted
to the FWHM using
$\mathrm{FWHM}=2\sqrt{2\ln 2}\sigma$ to estimate the effective resolution.
Figures S1(e) and S1(h) show the FWHM values for the individual bands; the
solid and dashed lines denote the mean and the mean $\pm$ standard deviation,
respectively. The resulting effective resolutions are
$(453.8 \pm 7.3)\,\mathrm{nm}$ for direct imaging without a grating and
$(140.0 \pm 8.7)\,\mathrm{nm}$ for rotating-grating reconstruction.

\vspace{1em}

\begin{center}
\includegraphics[width=\textwidth]{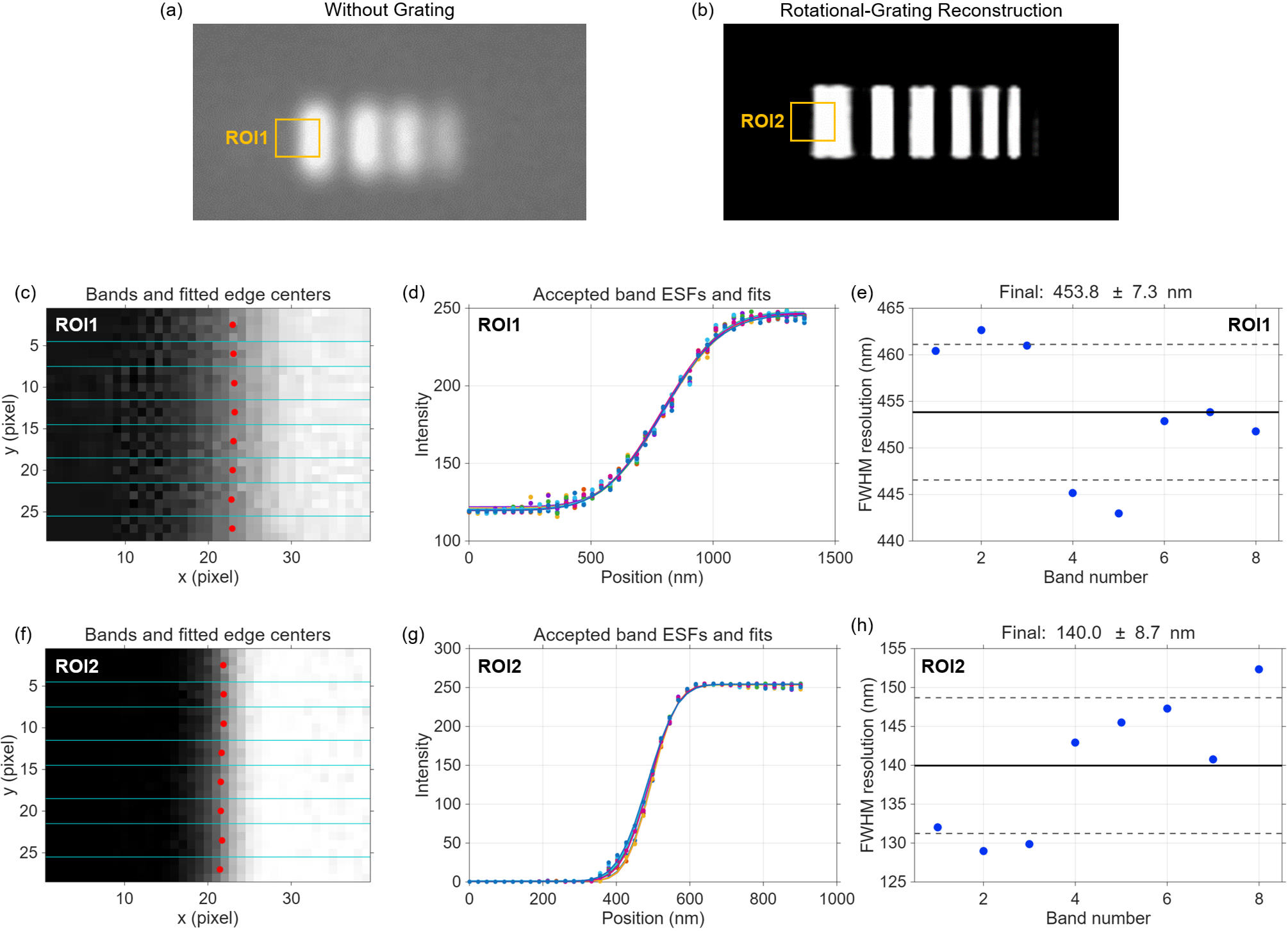}

\captionof{figure}{Estimation of effective lateral resolution from
stripe-edge responses.}
\label{fig:resolution}
\end{center}